\documentclass[a4paper]{jpconf}
\usepackage{amsmath,nicefrac,wasysym,aas_macros,graphicx,xcolor,hyperref}

\hypersetup{
  colorlinks=false,
  linkbordercolor= {white},
  citebordercolor = {white},
  urlbordercolor = {white}
}

\newcommand\cm{\,\rm cm}

\newcommand\g{\,\rm g}
\newcommand\erg{\,\rm erg}
\newcommand\K{\,\rm K}

\newcommand\au{\,\rm au}

\newcommand\perccm{\,\rm cm^{-3}}
\newcommand\Msun{\,M_\odot}

\newcommand\tms{\!\times\!}
\newcommand\cdt{\!\cdot\!}

\newcommand\bb{\hat{{\mathbf b}}}

\newcommand\V{\mathbf v}
\newcommand\B{\mathbf B}
\newcommand\F{\mathbf F_{\rm R}}

\newcommand\etao{\eta_{\rm O}}
\newcommand\etad{\eta_{\rm AD}}
\newcommand\EO{\mathbf E_{\rm O}}
\newcommand\EAD{\mathbf E_{\rm AD}}

\newcommand\eR{E_{\rm R}}
\newcommand\pR{\mathcal{P}_{\rm R}}
\newcommand\Qirr{\mathcal{Q}^+_{\rm irr}}
\newcommand\kP{\kappa_{\rm P}}
\newcommand\kR{\kappa_{\rm R}}
\newcommand\aR{a_{\rm R}}

\newcommand{\simgt}%
           {\,\hbox{\lower0.35ex\hbox{$\sim$}\llap{\raise0.35ex\hbox{$>$}}}\,}
\newcommand{\simlt}%
           {\,\hbox{\lower0.35ex\hbox{$\sim$}\llap{\raise0.35ex\hbox{$<$}}}\,}

\usepackage{color,ulem,soul}

\definecolor{dark-red}{rgb}{0.75, 0.00, 0.00}
\definecolor{hlcolor}{rgb}{1.00, 0.90, 0.85}
\sethlcolor{hlcolor}

\begin{document}

\title{Toward realistic simulations of magneto-thermal
       winds from weakly-ionized protoplanetary disks}

\author{Oliver Gressel}

\address{Niels Bohr International Academy, The Niels Bohr Institute,
  Blegdamsvej 17, DK-2100, Copenhagen \O, Denmark}

\ead{gressel@nbi.ku.dk}

\begin{abstract}
Protoplanetary disks (PPDs) accrete onto their central T~Tauri star
via magnetic stresses. When the effect of ambipolar diffusion (AD) is
included, and in the presence of a vertical magnetic field, the disk
remains laminar between 1-5$\,$au, and a magnetocentrifugal disk wind
forms that provides an important mechanism for removing angular
momentum. We present global MHD simulations of PPDs that include Ohmic
resistivity and AD, where the time-dependent gas-phase electron and
ion fractions are computed under FUV and X-ray ionization with a
simplified recombination chemistry. To investigate whether the mass
loading of the wind is potentially affected by the limited vertical
extent of our existing simulations, we attempt to develop a model of a
realistic disk atmosphere. To this end, by accounting for stellar
irradiation and diffuse reprocessing of radiation, we aim at improving
our models towards more realistic thermodynamic properties.
\end{abstract}


\section{Introduction}\vspace{2pt}

Interpreting observational properties of T~Tauri systems
\cite{2011ARA&A..49...67W} is closely tied to under\-standing the
complex dynamical evolution of gaseous protoplanetary disks (PPDs),
both in terms of their chemistry, and in terms of the microphysics
that govern the evolution of the embedded magnetic fields. Moreover,
with PPDs being the birth sites for extrasolar planets, a sound
physical picture of the dust \cite{2014prpl.conf..547J} and
planetesimal \cite{2012MNRAS.422.1140G} evolution is needed to
ultimately provide the building blocks for a comprehensive theory of
planet formation.

The fundamental drivers of disk evolution are mass loss processes and
redistribution of angular momentum \cite{2014prpl.conf..411T}. In
sufficiently ionized parts of the disk, the latter can be achieved by
turbulent stresses. When accounting for the ionization structure of a
typical PPD, large parts of the disk, however, remain laminar owing to
the effect of ambipolar diffusion (AD). In this situation, angular
momentum is primarily transported via a magnetocentrifugal disk wind
\cite{2013ApJ...769...76B,2015ApJ...801...84G}. While this picture is
further complicated when accounting for the Hall effect
\cite{2014A&A...566A..56L,2015MNRAS.454.1117S}, it has nevertheless
become clear that the thermal structure of the disk plays an important
role in setting the mass loading of the disk wind
\cite{2016ApJ...818..152B}, and hence the timescale on which the
system evolves \cite{2016ApJ...821...80B,2016arXiv160900437S}.


\section{Methods}\vspace{2pt}

As in our previous work \cite{2015ApJ...801...84G}, we are solving the
single-fluid MHD equations including Ohmic resistivity and ambipolar
diffusion, that is, the electromotive force resulting from the mutual
collision of ions and neutrals. The diffusion coefficients, $\etao$,
and $\etad$ are specified by means of a look-up table, which has been
obtained by solving a simple chemical ionization/recombination
network. The resulting electromotive forces stemming from the Ohmic
and ambipolar diffusion terms are then given by
\begin{equation}
  \EO \equiv - \etao \,(\nabla\tms\B)\,,\quad\textrm{and}
  \label{eq:emf_o}
\end{equation}
\begin{equation}
  \EAD \equiv \etad \,\left[\,(\nabla\tms\B)\times\bb\,\right]\times\bb\,,
  \label{eq:emf_ad}
\end{equation}
with $\bb\equiv \B/|\B|$ being the unit vector along $\B$, and where
the double vector product results in an additional minus sign, such
that positive coefficients $\etao$ and $\etad$ signify diffusion of
magnetic fields, with the latter only being sensitive to currents
perpendicular to the field direction.

For the typical number densities in the context of PPDs, the
frictional coupling between the ions and neutrals happens so quickly
(compared to dynamical timescales of interest) that we assume the
charged species move at their terminal velocity with respect to the
neutrals, that is, where the Lorentz force is balanced by the
drag. Because of the low degree of ionization, we formulate our
continuity equation and momentum conservation in terms of the neutral
component, although it does experience the Lorentz force as mediated
by particle collisions.

We use a modified version of the \textsc{nirvana-iii} finite volume
Godunov code \cite{2004JCoPh.196..393Z,2011JCoPh.230.1035Z}. The code
adopts a total-energy formulation\footnote{ Note that we do not
  include the radiation energy density, $\eR$, in the total energy (as
  it is sub-dominant in the problem we consider). We however retain
  the radiation flux in the momentum equation for consistency.}  with
conserved variables $\rho$, $\rho\V$, and $e\equiv \epsilon +
\rho\V^2\!/2 + \B^2\!/2$.  Together with the conservation of radiation
energy, $\eR$, and defining the total pressure, $p^{\star}$, as the
sum of the gas and magnetic pressures, the system of equations we
solve reads
\begin{eqnarray}
  \partial_t\rho +\nabla\cdt(\rho \V) & = & 0             \,, \\[4pt]
  \partial_t(\rho\V) +\nabla\cdt
     \big[\rho\mathbf{vv}+p^{\star}I-\mathbf{BB}\,\big] & = &
          - \rho \nabla\Phi + \rho\kR/c\,\F               \,, \\[4pt]
  \partial_t e + \nabla\cdt
     \big[(e + p^{\star})\V - (\V\cdt\B)\B\,\big] & = &   \label{eq:ene}
          \nabla \cdot \big[\, (\EO\!+\!\EAD)\tms\B\,\big]
        - \rho (\nabla\Phi)\cdt\V                    \nonumber\\
      & & +\,c\,\rho\kP\left(\eR-\aR\,T^4\right)
          + \rho\kR/c\,\F\cdt\V + \Qirr                   \,, \\[4pt]
  \partial_t \eR + \nabla\cdt (\eR\V) & = &               \label{eq:erad}
          -\,c\,\rho\kP\left(\eR-\aR\,T^4\right)
          - \nabla\cdt\F - \pR\!:\!\nabla\V               \,, \\[4pt]
  \partial_t \B -\!\nabla\tms\big[\, \V\tms\B             \label{eq:ind}
                           + \EO + \EAD \,\big] & = & 0   \,,
\end{eqnarray}
where $\F$ is the radiation flux, $\pR$ is the radiation pressure
tensor, $\kR$ and $\kP$ are the Rosseland and Planck mean opacities,
respectively, $\aR\equiv 4\, \sigma/c$ is the radiation density
constant, $T\equiv \bar{\mu}m_{\rm H}/k_{\rm B}\;p/\rho$ is the gas
temperature, and the gas pressure is $p = (\gamma-1)\epsilon$, where
$\gamma=7/5$ is chosen as appropriate for an ideal diatomic gas. The
gravitational term $\Phi(r) \equiv -G \Msun/r$ represents the
point-mass potential of the solar-mass star at the center of our
spherical-polar coordinate system, and $\Qirr$ represents external
irradiation heating due to the star.

\subsection{Flux limited diffusion}\smallskip

The above equations can only be solved once the radiation flux and
pressure tensor are specified. An attractive method for obtaining the
Eddington tensor that relates $\pR$ with $\eR$ is to solve for the
steady-state but angle-dependent radiation intensity and integrate the
first (for $\F$) and second (for $\pR$) order moments directly
\cite{2012ApJS..199...14J}. In the interest of maintaining minimum
algorithmic complexity and computational expense, we instead adopt the
classical \textit{ad hoc} closure
\begin{equation}
  \F = -\lambda(R)\,\frac{c}{\rho\kR}\nabla\,\eR\,,
  \label{eq:flux}
\end{equation}
that specifies $\F$ in terms of a diffusive flux with diffusion
coefficient $D\equiv \lambda(R) c /\rho\kR$, where
\begin{equation}
  R\equiv \frac{|\nabla\eR|}{\rho\kR\langle\eR\rangle}
\end{equation}
is a dimensionless number specifying how abruptly the radiation energy
density varies compared to the length scale defined by the optical
extinction coefficient $\rho\kR$, and where $\lambda(R)\rightarrow
1/R$ (for $R\gg 1$) is a limiter function \cite{1981ApJ...248..321L}
that guarantees $|\F|<c\eR$ in regions of low optical depth, that is,
in regions where the diffusion approximation is not valid. In the
optically thick limit, $\lambda(R)\rightarrow 1/3$ (for $R\ll 1$),
which corresponds to the Eddington approximation.

The described approach has its known shortcomings over
characteristics-based methods \cite{2012ApJS..199...14J}, but in
combination with stellar irradiation heating (determined using a
simplified ray tracing algorithm) it has been deemed an acceptable
compromise
\cite{2013A&A...549A.124B,2013A&A...555A...7K,2015A&A...574A..81R} in
terms of being able to incorporate radiation thermodynamics in fully
dynamical 3D simulations, whereas more accurate Monte Carlo methods
are comparatively expensive and have to deal with statistical noise
\cite{2012MNRAS.425.1430N}.

We currently treat both irradiation and diffuse redistribution in the
\textit{gray} approximation, but multigroup approaches are
straightforward \cite{2015A&A...578A..12G}, especially for the
irradiation component \cite{2013A&A...555A...7K,2015A&A...574A..81R},
wherein computational expenditure scales with the number of frequency
bins, and where increased realism can be achieved for the thermal
structure of the outer disk ($R\simgt 20\au$). For our PPD model, we
precompute look-up tables for mean opacities, $\kR(\rho,T)$ and
$\kP(\rho,T)$, using D.~Semenov's \texttt{opacity.f}
\cite{2003A&A...410..611S}, where a fixed dust-to-gas mass ratio of
$0.01395$ is assumed. To account for depletion of small dust by grain
growth, we enable the reduction of the obtained opacities by a scale
factor. Since the opacity tables combine contributions from dust
grains and gas molecules, this is only valid for temperatures below
the dust sublimation threshold.

\subsection{Reduced speed of light approximation}\smallskip

The thermodynamic coupling of the gaseous matter with the radiation
field is described by the $\,c\,\rho\kP\left(\eR-\aR\,T^4\right)$
terms in eqns.~(\ref{eq:ene}) and (\ref{eq:erad}),
respectively. Subsuming the $-\nabla\cdot\F$ term in
eqn.~(\ref{eq:erad}) with the definition of the diffusive flux
(\ref{eq:flux}) amounts to a diffusion equation for the redistribution
of radiation energy. Both effects can be combined into the subsystem
\begin{eqnarray}
  \partial_t \epsilon & = &
         +\,c\,\rho\kP\left(\eR-\aR\,T^4\right) \,, \\[4pt]
  \frac{c}{\hat{c}}\; \partial_t \eR & = &
          -\,c\,\rho\kP\left(\eR-\aR\,T^4\right)
          + \nabla\cdt\big[ D\, \nabla\eR \big]\,,  \label{eq:radif}
\end{eqnarray}
which we solve by means of operator splitting. Unlike in
ref.~\cite{2015A&A...574A..81R}, we do not include the
$\pR\!:\!\nabla\V$ term in this subsystem but instead treat it as a
regular source term when updating the main hyperbolic system of
equations. Since, for a large diffusion coefficient $D$, the parabolic
system (\ref{eq:radif}) becomes stiff, the most common approach is to
use implicit methods to solve it. Especially in view of applications
employing adaptive mesh refinement, we have chosen to avoid an
implicit update for $\nabla\cdot(D\, \nabla\eR)$, as it demands costly
non-local communication patterns, and a potentially expensive matrix
inversion.

Instead, we integrate the diffusion part of (\ref{eq:radif}) in a
time-explicit fashion, and use the \textit{reduced speed of light}
approach \cite{2001NewA....6..437G} to ameliorate the strict time-step
constraints. This method has recently been employed in the context of
simulations of the interstellar medium \cite{2013ApJS..206...21S}. The
approximation is valid as long as the radiative timescales resulting
from the adopted artificial value of $\hat{c}=\phi\,c$ (with
$\phi=\,$const.$\,<1$) are still short compared to any relevant
dynamical timescales. In the context of PPDs, we are mainly interested
in the role of radiative effects in setting the consistent temperature
structure of the disk, and true radiation hydrodynamic effects are
believed to only be of minor importance during the T~Tauri phase
\cite{1998apsf.book.....H}.

The approximation is introduced by amplifying the left-hand-side of
(\ref{eq:radif}) by a factor of $c/\hat{c}>\!1$. It is crucial that,
because all other terms remain unaffected, this implies the
modification does not alter the late-time steady-state solution, where
$\partial_t\rightarrow 0$, but only changes the \textit{timescale} on
which this solution is achieved. On practical grounds, we multiply
(\ref{eq:radif}) by $\hat{c}/c$, which implies replacing $c$ for
$\hat{c}$ in the radiation matter coupling, and attenuating the
diffusion term by a factor of $\phi$. This illustrates how the method
works to weaken the stiffness of the diffusion term. To explicitly
integrate (\ref{eq:radif}), we employ the second-order accurate
Runge-Kutta-Legendre (RKL2) super-time-stepping scheme
\cite{2012MNRAS.422.2102M}, that is already used for the updates of
the other parabolic terms (such as, viscosity, thermal conduction,
resistive and ambipolar diffusion) in the code.

\subsection{Radiation matter coupling}\smallskip
\label{sec:patankar}

Even with the reduced speed, $\hat{c}=\phi\,c$, the
radiation-matter-coupling term $\,\hat{c}\, \rho\kP
\left(\eR-\aR\,T^4\right)$ can itself contribute a severe timestep
constraint in regions of high opacity, where the coupling becomes
stiff. Ignoring, for the moment, the diffusion term in
(\ref{eq:radif}), the coupling amounts to an ordinary differential
equation, similar to production/destruction equations that are common
in other fields of science. While these are typically solved by
explicit Runge-Kutta (RK) methods, higher order predictor/corrector
schemes do not typically guarantee positivity or conservation of, in
our case, the energy $\bar{e}\equiv (\phi\,\epsilon+\eR)$.

There however exists a class of modified RK methods
\cite{BURCHARD20031} that employ the so-called Patankar trick -- an
implicit weighting of the production/destruction terms with the ratio
of the evolved quantity before and after the update. It can be shown
that for a single-step update, such a weighting precludes that
negative values are obtained. Moreover, since the weighting factors
are overall symmetric, conservation is guaranteed. Specifically, we
use the MPRK scheme given by eqn.~(27) in
ref.~\cite{BURCHARD20031}. The resulting update is formally implicit,
but can algebraically be manipulated into fully explicit form, that
is,
\begin{eqnarray}
  \Delta e & \equiv & \epsilon^{(n+1)} - \epsilon^{(n)} \nonumber\\
  & \equiv & \big(\eR^{(n)} - \eR^{(n+1)} \big) \phi^{-1} \; = \;
            \frac{\eR^{(n)} - \left(\aR\,T^4\right)^{(n)} }
               { \phi + \left(\aR\,T^4\right)^{(n)} / \epsilon^{(n)}
                 + \left(c\rho\kP \Delta t\right)^{-1} }\,,
\end{eqnarray}
for the predictor step, and
\begin{equation}
  \Delta e \; = \; \frac{ \eR^{\star}\,\epsilon^{\dagger} \eR^{(n)}
    - \left(\aR\,T^4\right)^{\star}\eR^{\dagger}\epsilon^{(n)} }
               { \phi \eR^{\star}\epsilon^{\dagger}
                 + \left(\aR\,T^4\right)^{\star}\eR^{\dagger}
                 + \eR^{\dagger}\epsilon^{\dagger}
                 \left(c\rho\kP \Delta t\right)^{-1} }\,,\\[4pt]
\end{equation}
for the corrector step, where $\epsilon^{\dagger}\equiv
\epsilon^{(n)}+\Delta e$ is the forward-Euler predictor value of
$\epsilon^{(n+1)}$, and $\epsilon^{\star}\equiv
0.5\,(\epsilon^{(n)}+\epsilon^{\dagger})$ is the time-averaged
state.\footnote{The same conventions, of course, apply for the terms
  $(\aR\,T^4)$, and $\eR$.} Compared to implicit methods, that demand
iterative root-finding, or so-called $\theta$-schemes (see, e.g.,
section~3.4 in ref.~\cite{2013ApJS..206...21S}) the method presented
here offers a relatively inexpensive, parameter-free non-iterative
alternative.

\subsection{Irradiation heating}\smallskip

Our existing global disk models
\cite{2013MNRAS.435.2610N,2013ApJ...779...59G} have either assumed a
locally-isothermal temperature $T=T(R)$, with $R$ being the
cylindrical radius from the star, or have used an adiabatic equation
of state with a Newtonian cooling term in the energy equation that
reinstated the $T(R)$ profile on a specified timescale (typically a
short fraction of the local orbital period).

Compared to this, even relatively basic models
\cite{1997ApJ...490..368C} of dust absorption and re-radiation of star
light in the disk surface, obtain a much more complex temperature
structure within the PPD -- with superheated surface layers and cool
interiors. To account for such effects, we include a
frequency-integrated radially attenuated irradiation flux
\begin{equation}
  \F\,\!_{,\rm irr}(r) = F(r_\star)\,\left(\frac{r_\star}{r}\right)^2\,
                         \exp\,(-\tau_{\rm P}(r))\; \hat{\mathbf{r}}
  \label{eq:firr}
\end{equation}
from the central star with effective temperature $T_\star=5780\K$, and
$r_\star=r_\odot$, and where the optical depth, $\tau_{\rm
  P}(r)\equiv\int_{r_\star}^{r}\,\rho(r')\,\kP(\rho,T_\star)\,{\rm
  d}r'$ is obtained by integrating along radial rays. Following
previous work
\cite{2013A&A...549A.124B,2013A&A...555A...7K,2015A&A...574A..81R}, we
obtain the irradiation energy source term $\Qirr$ by computing the
negative divergence of (\ref{eq:firr}) over each grid cell. In regions
of low optical depth, we use the integral formulation
\begin{equation}
  \Qirr(r_i) = \frac{\rho\kP}{\Delta V_r}
  \int_{r_{i-\nicefrac{1}{2}}}^{r_{i+\nicefrac{1}{2}}}
  \hat{\mathbf{r}}\cdt\F\,\!_{,\rm irr}(r')\,r'^2{\rm d}r'\,,
  \label{eq:Qirr}
\end{equation}
with $\Delta V_r=\nicefrac{1}{3}\,(r^3_{i+\nicefrac{1}{2}} \!-
r^3_{i-\nicefrac{1}{2}})$ instead, as this formulation has been found
to produce a more accurate solution on the discretized mesh, when
differences across cells are small \cite{1999A&A...348..233B}.


\section{Results}\vspace{2pt}

\subsection{Radiative transfer test problems}\smallskip

\begin{figure}[h]
  \includegraphics[width=0.6\columnwidth]{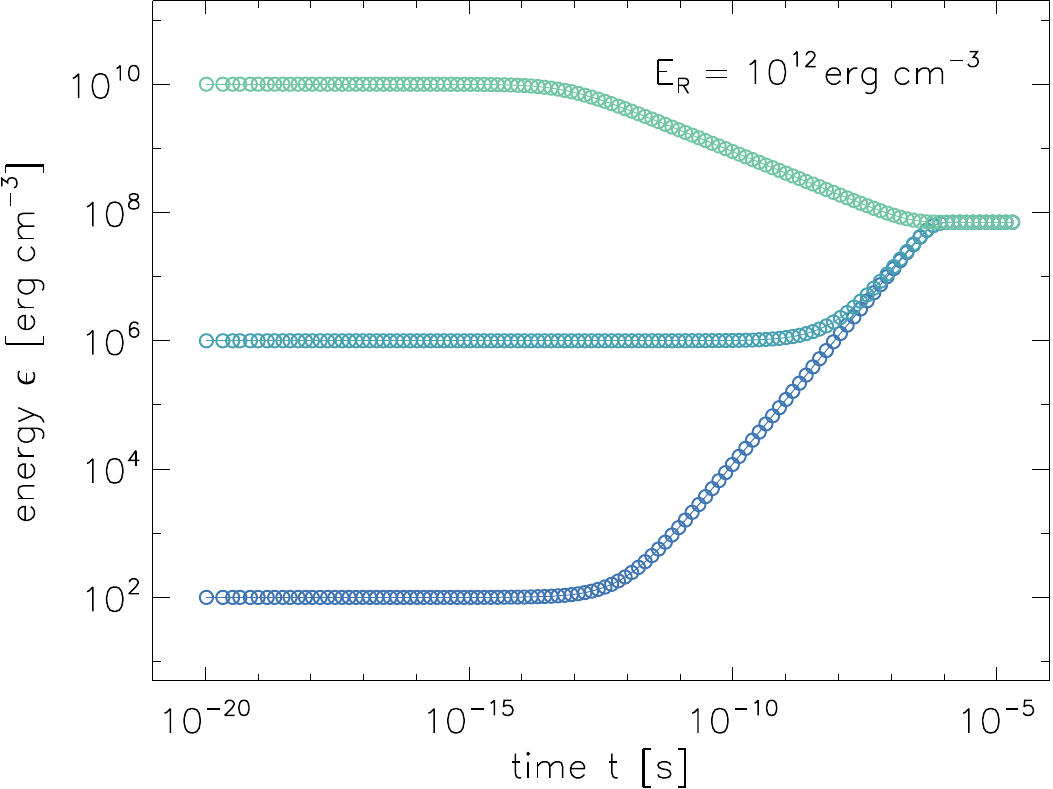}\hfill%
  \begin{minipage}[b]{0.27\columnwidth}%
    \caption{A simple single zone radiation matter coupling test, with
      $\kP=0.4\cm^2\g^{-1}$, $\bar{\mu}=0.6$,
      $\gamma=\nicefrac{5}{3}$, and
      $\rho=10^{-7}\g\perccm$. \label{fig:rm_coupling} }
  \end{minipage}
\end{figure}

To verify our implementation of the radiation-matter-coupling term, we
have performed a simple one-zone model \cite{2001ApJS..135...95T},
where the thermal energy density, $\epsilon$, is initially out of
balance with the radiation energy density $\eR=10^{12}\erg\perccm$. In
Fig.~\ref{fig:rm_coupling}, we show three cases with $\epsilon_0=
10^2$, $10^6$, and $10^{10}\erg\perccm$, which all converge to the
same final equilibrium state. For the purpose of plotting the curves,
we have limited the timestep artificially to sample timescales shorter
than the radiative equilibration timescale. We have, however, tested
that even for numerical time steps somewhat larger than the coupling
timescale the scheme remains stable, as expected from the
implicit-like integration scheme (see sect.~\ref{sec:patankar}) that
we use.

\begin{figure}[h]
  \includegraphics[width=0.75\columnwidth]{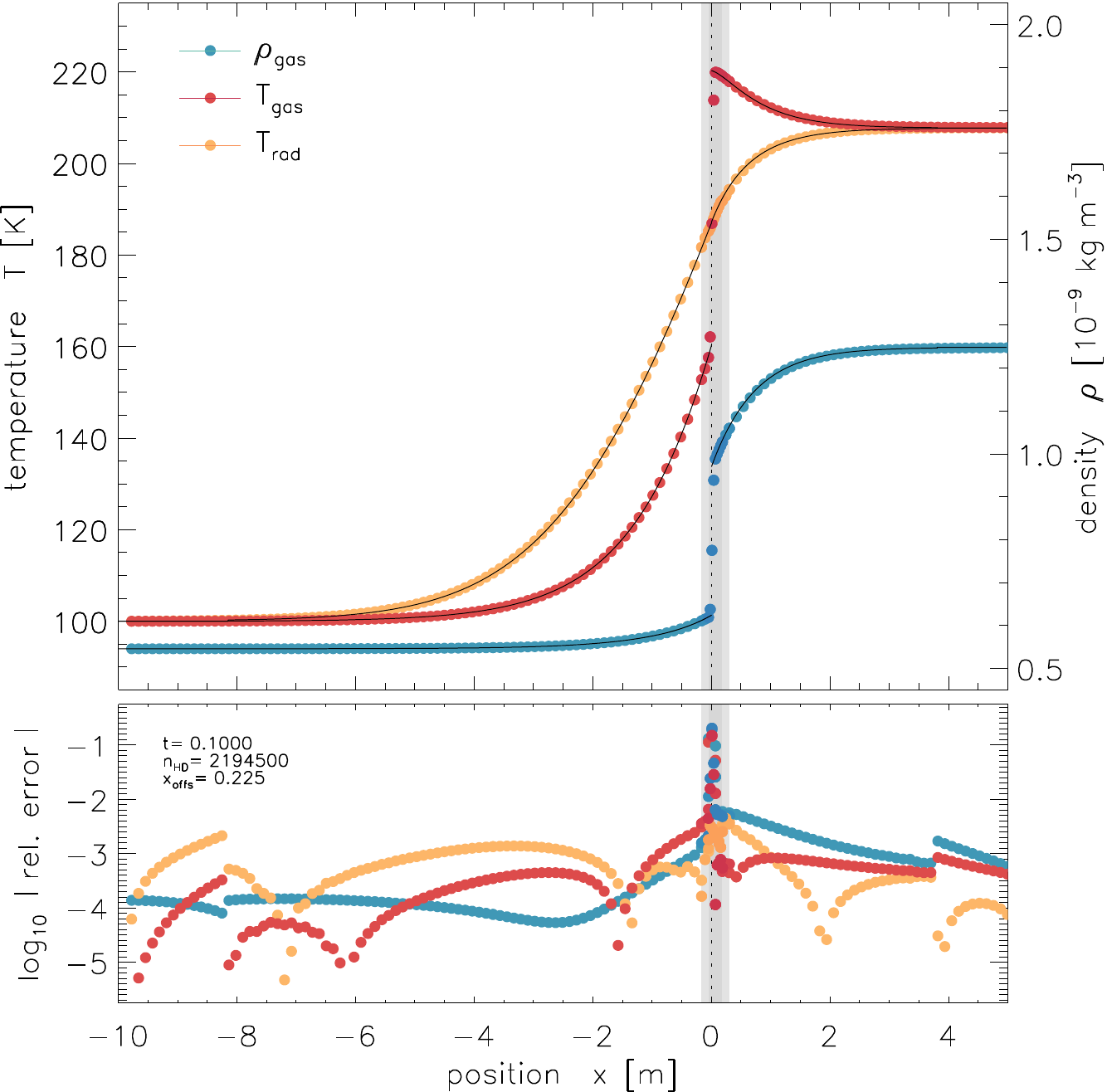}\hfill%
  \begin{minipage}[b]{0.25\columnwidth}%
    \caption{Radiative shock with ${\rm Ma}=2$ and resolution $n=128$
      +2 levels of adaptive mesh refinement (shaded
      areas). \emph{Upper panel:} Profiles of $\rho_{\rm gas}$
      (blue), $T_{\rm gas}$ (red), and $T_{\rm rad}$
      (yellow). \emph{Lower panel:} deviation from the semi-analytic
      solution (black). \label{fig:shock} }
  \end{minipage}
\end{figure}

A standard test case for assessing the interplay of the
radiation-matter-coupling with the radiation diffusion are radiative
shocks, for which there exist semi-analytic solutions in simplified
situations \cite{2008ShWav..18..129L}. In Fig.~\ref{fig:shock}, we
plot the solution of the ${\rm Ma}=2$ case from
ref.~\cite{2008ShWav..18..129L}, using $n=128$ grid cells in the
$x$~direction, as well as two levels of adaptive mesh refinement
(triggered by gradients in the thermal energy, $\epsilon$), which are
shown as gray shaded areas in the plot. As seen in the lower panel of
Fig.~\ref{fig:shock}, apart from the shock location (which has been
shifted by $x_{\rm offs}=22.5\cm$ with respect to the semi-analytic
solution), all quantities agree to within $\permil$ accuracy. We have
also successfully performed the harder ${\rm Ma}=5$ test, which we
omit here for brevity. An excellent description of the radiative shock
test, including the precise values used for initial conditions, can be
found in sect.~4.4 of ref.~\cite{2015A&A...574A..81R}.

\subsection{Preliminary global MHD simulations with irradiation}\smallskip

Returning to the original motivation for implementing radiative
physics in our modified version of the \textsc{nirvana-iii} code, we
conclude by presenting a preliminary snapshot of a global axisymmetric
protoplanetary disk simulation including Ohmic resistivity, ambipolar
diffusion, radiation diffusion, and stellar irradiation. The ultimate
goal of these simulations is to study how the mass-loading of the
magnetocentrifugal wind depends on the disk thermal structure.

\begin{figure}[h]
  \center\includegraphics[width=0.8\columnwidth]{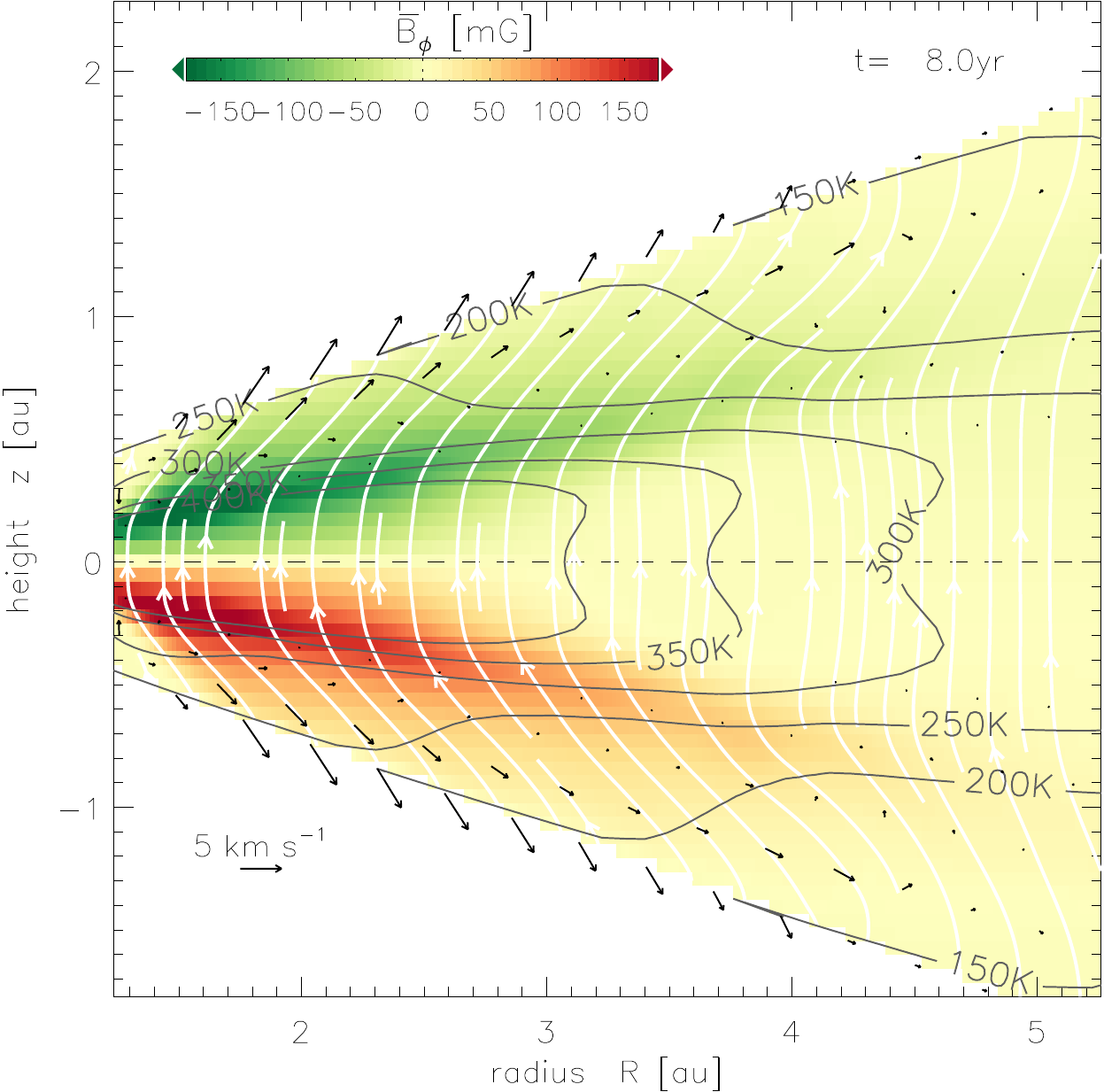}
  \caption{Detail from a proof-of-concept global MHD simulation of a
    PPD with ambipolar diffusion, radiative transfer and stellar
    irradiation, showing $\bar{B}_\phi$ (colour), poloidal velocity
    (black) and magnetic field lines (white), and iso-contours of the
    radiation temperature (grey). \label{fig:vis2D}}
\end{figure}

As a simple proof-of-concept, we present a close-up of the inner disk
in a simulation covering seven (initial) pressure scale heights in
latitude (see Fig.~\ref{fig:vis2D}). The basic disk setup is very
similar to the one used in ref.~\cite{2015ApJ...801...84G}, and we
have additionally used opacities that correspond to a dust-depletion
of a factor of ten compared to the typical interstellar
abundance. Similar to our previous simulations, the magnetic field
lines (white) bend outward, and in the upper disk layers, where the
matter is sufficiently coupled to the magnetic field, a
magnetocentrifugal disk wind ensues (black vectors).

Iso-contour lines (gray) of the radiation temperature $T_{\rm rad}
\!\equiv\! \eR^{\nicefrac{1}{4}}\aR^{\nicefrac{-1}{4}}$ illustrate the
disk's thermal structure that deviates noticeably from the
constant-on-cylinders radial temperature profile $T=T(R)$, that we
have assumed previously. The presented preliminary run used a
reduced-speed-of-light factor $\phi=10^{-4}$. Further tests will show
whether the chosen time-explicit framework is powerful and efficient
enough to be of use in realistic situations.


\ack The author thanks Jon Ramsey for many useful discussions, for
carefully reading this manuscript, and for providing his code to
compute semi-analytic reference solutions of radiative shocks. Dmitry
Semenov is acknowledged for providing helpful clues in connection with
his opacity code. The research leading to these results has received
funding from the European Research Council (ERC) under the European
Union's Horizon 2020 research and innovation programme (grant
agreement No 638596).


\section*{References}\vspace{2pt}


\begin{thebibliography}{10}
\expandafter\ifx\csname url\endcsname\relax
  \def\url#1{{\tt #1}}\fi
\expandafter\ifx\csname urlprefix\endcsname\relax\def\urlprefix{URL }\fi
\providecommand{\eprint}[2][]{\url{#2}}

\bibitem{2011ARA&A..49...67W}
{Williams} J~P and {Cieza} L~A 2011 {\em \araa\/} {\bf 49} 67--117
  (\textit{Preprint} \eprint{1103.0556})

\bibitem{2014prpl.conf..547J}
{Johansen} A, {Blum} J, {Tanaka} H, {Ormel} C, {Bizzarro} M and {Rickman} H
  2014 {\em PPVI\/}  547

\bibitem{2012MNRAS.422.1140G}
{Gressel} O, {Nelson} R~P and {Turner} N~J 2012 {\em \mnras\/} {\bf 422}
  1140--1159 (\textit{Preprint} \eprint{1202.0771})

\bibitem{2014prpl.conf..411T}
{Turner} N~J, {Fromang} S, {Gammie} C, {Klahr} H, {Lesur} G, {Wardle} M and
  {Bai} X~N 2014 {\em PPVI\/}  411--432

\bibitem{2013ApJ...769...76B}
{Bai} X~N and {Stone} J~M 2013 {\em \apj\/} {\bf 769} 76 (\textit{Preprint}
  \eprint{1301.0318})

\bibitem{2015ApJ...801...84G}
{Gressel} O, {Turner} N~J, {Nelson} R~P and {McNally} C~P 2015 {\em \apj\/}
  {\bf 801} 84 (\textit{Preprint} \eprint{1501.05431})

\bibitem{2014A&A...566A..56L}
{Lesur} G, {Kunz} M~W and {Fromang} S 2014 {\em \aap\/} {\bf 566} A56
  (\textit{Preprint} \eprint{1402.4133})

\bibitem{2015MNRAS.454.1117S}
{Simon} J~B, {Lesur} G, {Kunz} M~W and {Armitage} P~J 2015 {\em \mnras\/} {\bf
  454} 1117--1131 (\textit{Preprint} \eprint{1508.00904})

\bibitem{2016ApJ...818..152B}
{Bai} X~N, {Ye} J, {Goodman} J and {Yuan} F 2016 {\em \apj\/} {\bf 818} 152
  (\textit{Preprint} \eprint{1511.06769})

\bibitem{2016ApJ...821...80B}
{Bai} X~N 2016 {\em \apj\/} {\bf 821} 80 (\textit{Preprint}
  \eprint{1603.00484})

\bibitem{2016arXiv160900437S}
{Suzuki} T~K, {Ogihara} M, {Morbidelli} A, {Crida} A and {Guillot} T 2016 {\em
  A\&A\/} (\textit{Preprint} \eprint{1609.00437})

\bibitem{2004JCoPh.196..393Z}
{Ziegler} U 2004 {\em \rm JCoPh\/} {\bf 196} 393--416

\bibitem{2011JCoPh.230.1035Z}
{Ziegler} U 2011 {\em \rm JCoPh\/} {\bf 230} 1035--1063

\bibitem{2012ApJS..199...14J}
{Jiang} Y~F, {Stone} J~M and {Davis} S~W 2012 {\em \apjs\/} {\bf 199} 14
  (\textit{Preprint} \eprint{1201.2223})

\bibitem{1981ApJ...248..321L}
{Levermore} C~D and {Pomraning} G~C 1981 {\em \apj\/} {\bf 248} 321--334

\bibitem{2013A&A...549A.124B}
{Bitsch} B, {Crida} A, {Morbidelli} A, {Kley} W and {Dobbs-Dixon} I 2013 {\em
  \aap\/} {\bf 549} A124 (\textit{Preprint} \eprint{1211.6345})

\bibitem{2013A&A...555A...7K}
{Kuiper} R and {Klessen} R~S 2013 {\em \aap\/} {\bf 555} A7 (\textit{Preprint}
  \eprint{1305.2197})

\bibitem{2015A&A...574A..81R}
{Ramsey} J~P and {Dullemond} C~P 2015 {\em \aap\/} {\bf 574} A81
  (\textit{Preprint} \eprint{1409.3011})

\bibitem{2012MNRAS.425.1430N}
{Noebauer} U, {Sim} S, {Kromer} M, {R{\"o}pke} F and {Hillebrandt} W 2012 {\em
  \mnras\/} {\bf 425} 1430 (\textit{Preprint} \eprint{1206.6263})

\bibitem{2015A&A...578A..12G}
{Gonz{\'a}lez} M, {Vaytet} N, {Commer{\c c}on} B and {Masson} J 2015 {\em
  \aap\/} {\bf 578} A12 (\textit{Preprint} \eprint{1504.01894})

\bibitem{2003A&A...410..611S}
{Semenov} D, {Henning} T, {Helling} C, {Ilgner} M and {Sedlmayr} E 2003 {\em
  \aap\/} {\bf 410} 611--621

\bibitem{2001NewA....6..437G}
{Gnedin} N~Y and {Abel} T 2001 {\em NewA\/} {\bf 6} 437--455 (\textit{Preprint}
  \eprint{astro-ph/0106278})

\bibitem{2013ApJS..206...21S}
{Skinner} M~A and {Ostriker} E~C 2013 {\em \apjs\/} {\bf 206} 21
  (\textit{Preprint} \eprint{1306.0010})

\bibitem{1998apsf.book.....H}
{Hartmann} L 1998 {\em {Accretion Processes in Star Formation}\/}

\bibitem{2012MNRAS.422.2102M}
{Meyer} C~D, {Balsara} D~S and {Aslam} T~D 2012 {\em \mnras\/} {\bf 422}
  2102--2115

\bibitem{BURCHARD20031}
Burchard H, Deleersnijder E and Meister A 2003 {\em AppliedNumMath\/} {\bf 47}
  1 -- 30 ISSN 0168-9274

\bibitem{2013MNRAS.435.2610N}
{Nelson} R~P, {Gressel} O and {Umurhan} O~M 2013 {\em \mnras\/} {\bf 435}
  2610--2632 (\textit{Preprint} \eprint{1209.2753})

\bibitem{2013ApJ...779...59G}
{Gressel} O, {Nelson} R~P, {Turner} N~J and {Ziegler} U 2013 {\em \apj\/} {\bf
  779} 59 (\textit{Preprint} \eprint{1309.2871})

\bibitem{1997ApJ...490..368C}
{Chiang} E~I and {Goldreich} P 1997 {\em \apj\/} {\bf 490} 368--376
  (\textit{Preprint} \eprint{astro-ph/9706042})

\bibitem{1999A&A...348..233B}
{Bruls} J~H~M~J, {Vollm{\"o}ller} P and {Sch{\"u}ssler} M 1999 {\em \aap\/}
  {\bf 348} 233--248

\bibitem{2001ApJS..135...95T}
{Turner} N~J and {Stone} J~M 2001 {\em \apjs\/} {\bf 135} 95--107
  (\textit{Preprint} \eprint{astro-ph/0102145})

\bibitem{2008ShWav..18..129L}
{Lowrie} R~B and {Edwards} J~D 2008 {\em Shock Waves\/} {\bf 18} 129--143

\end{thebibliography}

\providecommand{\newblock}{}

\end{document}